\documentclass[useAMS,usenatbib]{mn2e}

\usepackage[english]{babel}
\usepackage{graphicx}
\usepackage{txfonts}
\usepackage{color}
\usepackage{aas_macros}

\title[{\it CoRoT} multicolour photometry of exoplanets]{An analysis of {\it CoRoT} multicolour photometry of exoplanets
\thanks{The {\it CoRoT} space mission was developed and  is operated by the French
space agency CNES, with participation of ESA's RSSD and Science Programmes,
Austria, Belgium, Brazil, Germany, and Spain.}}
\author[F. Borsa and E. Poretti]{F. Borsa$^{1,2}$\textsuperscript{\thanks{E-mail: francesco.borsa@brera.inaf.it}} 
 and E.
Poretti$^{2}$\\
$^{1}$Universit\`a dell'Insubria - Dipartimento di Scienza e Alta Tecnologia, Via Valleggio 11, 22100 Como, Italy\\
$^{2}$INAF -- Osservatorio Astronomico di Brera,
              Via E. Bianchi 46, 23807 Merate (LC), Italy}
\begin{document}

\date{Accepted 2012 September 26. Received 2012 September 17; in original form 2012 July 13}

\pagerange{\pageref{firstpage}--\pageref{lastpage}} \pubyear{2012}

\maketitle

\label{firstpage}

\begin{abstract}
We analysed the chromatic data of the planetary transits observed with {\it CoRoT}
to supply homogeneous time series in each of the {\it CoRoT} colours.
In a first step, we cleaned  the chromatic light curves from the
contamination of nearby stars 
and removed outliers and trends caused by
anything different from the planetary transits. 
Then, we simultaneously fitted  the chromatic transits of each planet, obtaining
a common solution for the orbital parameters $i$, $t_0$ and $a/R_s$,   with 
a particular care in the fitting for 
different limb-darkening parameters. The 
 planet-to-star radius ratios in the {\it CoRoT} colours are compatible 
when considering one planet at a time, but the ensemble of
low-mass planets seems to show a peculiar behaviour of the radius ratios.
\end{abstract}


\section{Introduction}

The discovery of new planets is recurrent news in the exploitation of
ground--based surveys (e.g., Mayor et al. 2011) and space missions (e.g., Borucki et al. 2011).
We count more than 700 confirmed detections and a lot of candidates are in the waiting list.
Therefore, it is quite natural to direct our attention
to techniques that can help in disentangling the uncertain cases and/or supply
more information on the planet's structure.
If we are able to point out any kind of new feature, this can help
in planning future investigations and/or theoretical analysis.

{\it Kepler} and {\it CoRoT} (convection, rotation and planetary transits; Baglin et al. 2006) 
space missions make use of the photometric method of transits to detect exoplanets.
Theory  predicts that the depth of the transit of a planetary body in front of its parent star should not depend on the
wavelength, except for the effect of the stellar limb darkening.
However this is true only at first order: the apparent radius of the planet, 
and therefore its transit depth, does depend on
the wavelength; the light from the parent star filters through the planet's atmosphere, 
so that the atmospheric signatures
remain imprinted in the transit depths at different wavelengths, creating a transmission spectrum.
The weight of this effect depends on the atmospheric scale height. After the first
{\it Hubble Space Telescope}  positive result in 2002, the detection
of spectral features in the exoplanet atmospheres over a broad wavelength range 
is now a firmly established field (Seager \& Deming 2010).

In the case of the {\it CoRoT}
mission, a chromatic device was mounted
in the focal block with the aim of discriminating `nearly
achromatic planetary transits from coloured stellar fluctuations' (Barge et al. 2006).
The light passes through a prism before entering in the exoplanet channel:
to obtain chromatic light curves, the mask is divided into three submasks
along the dispersion direction.
The mask (and so the relative submasks) for each target has a shape optimized for the magnitude and location
of nearby fainter stars. Such a shape is
chosen between a set of templates  in order to maximize the signal-to-noise ratio (S/N).
The chromatic device plays an important role in the selection of the most promising targets,
allowing the rejection of several false positives (Carpano et al. 2009), and the 
confirmation of the planetary nature of objects like CoRoT-7b (L{\'e}ger et al. 2009)
and CoRoT-8b (Bord{\'e} et al. 2010).

The specificity of each mask constitutes a great difficulty in extracting physical 
parameters from the coloured light curves, but the efforts involved could be rewarding.
Indeed, the detection of  small differences  in the amplitudes and in the phases
of the chromatic light curves of the high-amplitude $\delta$~Sct star CoRoT 101155310 
allowed us to discriminate between the radial and non-radial nature of the pulsation mode
(Poretti et al. 2011).
Therefore, we considered it noteworthy to undertake a study of the chromatic light curves of the 
first planets discovered by {\it CoRoT} to  point out  the useful hints that 
this unique photometric device can supply for further investigations.

\vspace{-1em}
\section{Data Analysis}

At the time of our analysis, {\it CoRoT} has discovered 23 planets, of which only the first 15 have public data.
Among these 15 planets, only 10 belong 
to the subset for which chromatic data are available: our analysis focused on these targets (Table \ref{contaminants}).

\subsection{Removing contamination}

\begin{table*} 
\centering
\caption{Estimates of the contamination factors in each mask (white) and submask (red, green, blue)
for the 10 {\it CoRoT} planets analysed, and characteristics of the parent stars. Nearby stars have been considered up 
to a 20~arcsec distance from the source. 
}
\setlength{\tabcolsep}{6pt}
\begin{tabular}{ccccc c cccc}
\hline
Planet & \multicolumn{4}{c}{{\it CoRoT} colour contamination}& Parent & $V$ & \multicolumn{2}{c}{Contaminants} & Discovery \\
 & White & Red & Green & Blue& Star & mag & Num. & $V$ mag range & paper \\
 & (per cent) &  (per cent) &  (per cent) &  (per cent) & &&&& \\
\hline
\noalign{\smallskip}
\smallskip
CoRoT-1b  & $1.02^{+0.31}_{-0.23}$ & $1.43^{+0.44}_{-0.32}$ & $0.36^{+0.14}_{-0.11}$ & $0.16^{+0.07}_{-0.06}$ & G0\,V & 13.62 & 6 & 17.25 - 20.50 & Barge et al. (2008) \\
\smallskip
CoRoT-2b  & $5.09^{+0.60}_{-0.53}$ & $2.94^{+0.58}_{-0.52} $&$ 6.41^{+1.27}_{-1.14}$ & $11.95^{+1.72}_{-1.53}$& G7\,V & 12.57 & 9 & 15.60 - 20.28 & Alonso et al. (2008) \\
\smallskip
CoRoT-4b  & $0.01^{+0.00}_{-0.00}$ & $0.01^{+0.00}_{-0.00} $&$ 0.00^{+0.00}_{-0.00}$ &$ 0.00^{+0.00}_{-0.00}$&F8\,V & 13.69 & 6 & 17.84 - 20.38 & Aigrain et al. (2008) \\
\smallskip
CoRoT-5b  & $3.28^{+0.90}_{-0.61}$ & $2.59^{+0.87}_{-0.65} $&$ 1.55^{+0.50}_{-0.33}$ &$ 6.38^{+1.50}_{-1.18}$&F9\,V & 14.02 & 6 & 16.19 - 20.32 & Rauer et al. (2009) \\
\smallskip
CoRoT-6b  & $0.96^{+0.24}_{-0.16}$ & $1.12^{+0.32}_{-0.21} $&$ 0.42^{+0.11}_{-0.09}$ & $0.60^{+0.22}_{-0.17}$&F9\,V & 13.91 & 6 & 16.59 - 20.58 & Fridlund et al. (2010)\\
\smallskip
CoRoT-7b  & $0.71^{+0.19}_{-0.19}$ & $0.08^{+0.03}_{-0.02} $&$ 0.08^{+0.02}_{-0.02}$ & $3.77^{+1.11}_{-1.18}$&K0\,V & 11.67 & 6 & 13.65 - 18.96 & Leger et al. (2009) \\
\smallskip
CoRoT-8b  & $0.96^{+0.24}_{-0.20}$ & $0.71^{+0.26}_{-0.19} $&$ 0.22^{+0.07}_{-0.05}$ & $2.79^{+0.83}_{-0.91}$&K1\,V & 14.30 & 6 & 18.30 - 20.10 & Bord{\'e} et al. (2010) \\
\smallskip
CoRoT-9b  & $0.56^{+0.13}_{-0.13}$ & $0.08^{+0.02}_{-0.01} $&$ 0.02^{+0.02}_{-0.01}$ & $3.01^{+0.74}_{-0.70}$&G3\,V & 13.69 & 6 & 15.04 - 20.68 & Deeg et al. (2010) \\
\smallskip
CoRoT-11b  & $4.00^{+0.65}_{-0.55}$ & $5.17^{+0.99}_{-0.82} $&$ 1.08^{+0.37}_{-0.23}$ & $1.90^{+0.44}_{-0.34}$&F6\,V & 12.94 & 7 & 16.35 - 20.60 & Gandolfi et al. (2010)\\

\noalign{\smallskip}

CoRoT-3b  & $7.02^{+0.97}_{-0.84}$ & $2.19^{+0.45}_{-0.41}$ & $11.14^{+1.99}_{-1.84}$ & $17.60^{+3.49}_{-2.74}$&F3\,V & 13.29 & 9 & 14.60 - 21.13 & Deleuil et al. (2008)\\

\noalign{\smallskip}
\hline
\end{tabular}
\label{contaminants}
\vspace{-2em}
\end{table*}

The point spread function (PSF) of the {\it CoRoT} optics is highly defocused, to improve the photometric precision and to reduce 
the sensitivity to satellite jitter. Such defocusing increases the probability that the PSF of the
star harbouring the planet overlaps the PSFs of nearby stars.
None of the N2 coloured light curves is corrected for the fluxes of stars entering the target star mask.
There is only an estimate of the contamination factors made with a generic mask before the run,
but only for the white light curve. An analysis of the contamination effects in the {\it CoRoT} colours
was performed in the case of CoRoT-8b only (Bord{\'e} et al. 2010).
Since the quantitative evaluation of the contamination effect is necessary to compare transit depths in different colours,
we applied a similar procedure by using the data available in the EXODAT database (Deleuil et al. 2009).

\begin{figure}
\centering
\includegraphics[width=8.5cm]{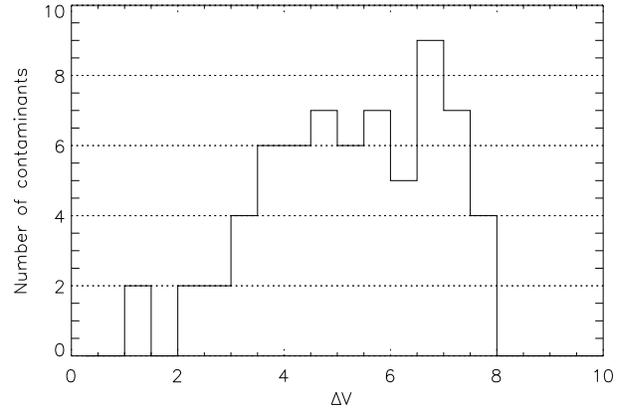}
\caption{Histogram representing the $V$ magnitude differences between the main targets
and the contaminant stars. Bins refer to intervals of 0.5~mag.}
\label{histocont}
\end{figure}

A preliminary step was the evaluation of the projected image on the CCD of the contaminant stars only, 
without the main target.
We calculated the inflight PSF for each target and 
then we applied it to the nearest six contaminant stars (i.e., those listed in EXODAT),
scaling the calculated PSF to match their $V$ magnitudes and $B-V$ colour indices. 
If the six EXODAT contaminants  were all closer than 20~arcsec
to the target, we completed the list with stars taken from the USNO-A2 catalogue, in order to be sure
to include all the potential contaminating sources. 
Fig. \ref{histocont} shows the histogram of the magnitude differences $\Delta V$ between the 67 contaminants and
the stars harbouring planets. 
In two cases only we have $\Delta V<2.0$ mag, related  
to contaminants far away from the main targets, i.e. 17.2 and 18.7~arcsec. This is not
surprising, since the masks are actually chosen to minimize the contamination effect. However,
although very small, this effect cannot be neglected for our purposes. 

The above procedure supplied us a clear starting point, i.e. it allowed us to evaluate how many pixels
of the target mask were contaminated by the flux from other stars, on the basis of their position,
magnitude and colour. To proceed in the photometric analysis, 
we had to take into account not only the different colours of the stars but also
the impossibility of having accurate photometric transformations between the
$V$ and $B-V$ values  and the spectral response of the specific mask+star combination.

To tackle this severe, intrinsic problem of the {\it CoRoT} photometry, we divided 
the calculated PSF of each contaminant into two parts, one
bluer and one redder. We scaled the bluer part as the $B$ mag and the redder part 
as the $V$ mag with respect to the main target.

Then, we integrated the flux of the contaminants in the main mask and in the three submasks, 
comparing it with the flux produced by the main target.
To estimate the contamination factor and its error bar we repeated the computations by running the
procedure 20 000 times for each star.  
We left the parameters vary in wide ranges: 
the $B$ and $V$ magnitudes by $\pm$0.3~mag (and hence the $B-V$ colour by $\pm$0.6~mag)
and the center of the image of each star in a $3\times3$~pixel$^2$ box.
The choice of these wide ranges was done for several reasons. EXODAT does not provide information 
about uncertainties on both magnitude and position of the contaminants, which are very faint stars. 
Moreover, the large 
error bar on the colour information is a way to take into account the impossibility of using
the exact {\it CoRoT} spectral response on contaminants. 
The magnitude range also accounts for possible light variability.
At the end of the process we chose the median values as the best estimate 
of the contamination factors and their error 
as the 68 per cent confidence interval (Table~\ref{contaminants}).
The contamination factors for CoRoT-8b are in excellent agreement with those estimated by Bord{\'e} et al. (2010),
i.e. 0.9, 0.7, 0.2, and 2.4 per cent for the white, red, green and blue {\it CoRoT} colours 
(hereafter $W_C$, $R_C$, $G_C$, $B_C$), respectively. This agreement strengthened our confidence in the adopted
procedure.

The green submask is always composed of only one column 
of pixels: it acts as a separator between the blue and red spectra.
As a consequence, the green flux is low compared to the others, and suffers from a higher dispersion. 
Hence we decided to disregard the $G_C$ light curve, concentrating
only on the $R_C$ and $B_C$ channels.

\subsection{Cleaning and detrending light curves}

After having corrected each light curve with the contamination factor estimates,
we analysed all the light curves separately, because trends and outliers are different in each colour.

We did not consider observations acquired when the satellite crossed the South Atlantic Anomaly and those
flagged as inaccurate by the pipeline. After that, we analysed the {\it CoRoT} time series to study each  
planetary transit. The light curves are not linear: they show trends due to the ageing of detectors
(Auvergne et al. 2009) and jumps due to hot pixels hit by cosmic rays.
After the removal of obvious outliers,
we calculated the average flux difference between two consecutive points.
We deleted the points which were more than three times this quantity both from the previous 
and the next points.

\begin{figure} 
\begin{center}
\includegraphics[width=8.5cm]{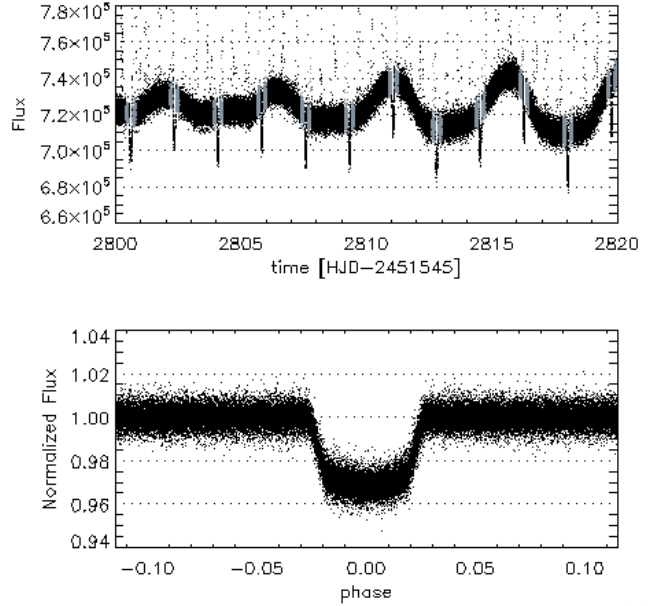}
\caption{Example of how our algorithm works. Top: raw white light curve of CoRoT-2b, in grey are evidenced
the data points we considered for the second order fit before and after each transit. Bottom: the final phase-folded transit
of CoRoT-2b after the correction algorithm.
}
\label{spianamento}
\end{center}
\end{figure}

Several  algorithms have been proposed to detrend {\it CoRoT} light curves (e.g. Mazeh et al. 2009).
We applied a specific approach for our purpose, i.e. having well-cleaned curves around the observed transits. 
We selected two intervals in the out-of-transit part of the light curves, 
before and after the central time of each transit $T_i$,
where the light curve is expected to be flat, but actually is affected by stellar activity and 
instrumental trends (Fig.~\ref{spianamento}, top panel).
As a rule,
we considered the intervals [$T_i-2.0~D,T_i-0.7~D]$ and [$T_i+0.7D,T_i+2.0~D]$, where $D$ is the full duration
of the transit. The data in these two intervals were fitted by a second-order polynomial. Then, we considered
all the data in the interval [$T_i-2.0~D,T_i+2.0~D]$, and we divided them by the second-order fit, thus removing
the trend and normalizing the flux units to the out-of-transit level (Fig.~\ref{spianamento}, bottom panel).
In such a way, we also corrected for stellar activity.
The calculated errors on all the detrending parameters have been
propagated through the entire analysis; the errors on the orbital
periods cited in the literature have also been taken into account when
phase-folding the transits.
We then superposed all the transits of each light curve using the periods determined in the discovery
papers (Table~\ref{references}) to obtain the phase-folded transits in each colour.

\begin{table} 
\centering
\caption{List of the {\it CoRoT} planets analysed, with
relative periods, ephemerides and masses taken from the discovery papers listed in Table~\ref{contaminants}.
}
\setlength{\tabcolsep}{5pt}
\begin{tabular}{cccc}
\hline
Planet & Period $(d)$ & $t_0-245 0000$ & Mass  $(M_J)$ \\
\hline
\noalign{\smallskip}

CoRoT-1b  & $1.5089557\pm0.0000064$ &$4159.4532$ & 1.03\\

CoRoT-2b   & $1.74299641\pm0.0000017$ &$4706.4041$ &  3.31\\

CoRoT-4b  & $9.20205\pm0.00037$ &$4141.36416$ &  0.72\\

CoRoT-5b   & $4.0378962\pm0.0000019$ &$4400.19885$ &  0.467\\

CoRoT-6b  & $8.886593\pm0.000004$ &$4702.2556$ & 2.96 \\

CoRoT-7b  & $0.853585\pm0.000024$ &$4398.0767$  & 0.015\\

CoRoT-8b  & $6.21229\pm0.00003$ &$4238.9743$ & 0.22 \\

CoRoT-9b  & $95.2738\pm0.0014$ &$4603.3447$ & 0.84 \\

CoRoT-11b & $2.99433\pm0.000011$ & $4597.679$  & 2.33\\

\noalign{\smallskip}

CoRoT-3b  & $4.25695\pm0.000005$ & $4283.1383$ & 21.66\\
\noalign{\smallskip}
\hline
\end{tabular}
\label{references}
\end{table}

\subsection{Limb darkening and transit fitting}\label{limb}

The treatment of limb darkening in the fitting of planetary transits has long been analysed.
Strong efforts were made to calculate the theoretical limb-darkening parameters, depending on the characteristics of the 
parent star and on the bandpass of the measurements (e.g. Claret 1998). With the coming of the  
{\it CoRoT} and {\it Kepler} space-based observatories, new effort was necessary to calculate new parameters for their 
ranges of sensitivity (Sing 2010; Claret \& Bloemen 2011).
However, no estimation has ever been made for the {\it CoRoT} colours. The {\it CoRoT} colours in fact are not bandpasses, 
because they are split by a low-dispersion device, and so the exact range of wavelengths slightly 
varies for each star, depending on its colour and  position on the CCD; this lack of a precise photometric system is preventing the
 theoretical estimation of the limb-darkening coefficients.
Howarth (2011) stresses how difficult it is to use standard and/or theoretical limb-darkening parameters when
studying exoplanetary systems.
We then decided to assume the general quadratic law for the limb-darkening effect only and 
to compute the $\mu_1$ and $\mu_2$ coefficients as free parameters in each case.

The phase-folded coloured light curves were fitted with the {\sc tap} package (Gazak et al. 2011), which uses the 
Mandel \& Agol (2002) model for the transit shape.
We performed a Markov Chain Monte Carlo analysis of 1000 000 steps for each planet, fitting simultaneously 
the 3 light curves ($W_C$, $R_C$, $B_C$) keeping the orbital period fixed. 
The inclination angle $i$, the distance between the
planet and the star $a$ (in units of the stellar radius $R_s$) and the transit timing $t_0$
were calculated considering all the colours simultaneously. 
On the other hand,
the planet-to-star radius ratios $R_p/R_s$ (the square root of the transit depth) and the limb-darkening 
coefficients $\mu_1$ and $\mu_2$ were calculated for each colour (Table \ref{depths}).
The limb-darkening parameters tabulated by Sing (2010) for the {\it CoRoT} range of sensitivity
are  within the error bars of our values for the $W_C$ light curves (Table~\ref{ldwhite}), thus corroborating our methodological approach.
The reported error bars are calculated by propagating the errors on the parameters 
(the long-term trend in different colours, the orbital periods and the contamination effects)
we used to extract the shape of the transit from the light curves.

\begin{table*} 
\centering
\caption{Fit results for the phase-folded transits of the planets analysed, with planet-to-star radius ratios and 
limb-darkening coefficients (linear and quadratic) for the $R_C$ and $B_C$ colours. The orbital periods are taken from the 
discovery papers.}
\setlength{\tabcolsep}{5.5pt}
\begin{tabular}{cccccccccrr}
\hline
Planet & $a/R_s$ & $i$ & \multicolumn{2}{c}{$R_p/R_s$} && \multicolumn{2}{c}{$\mu_1$} && \multicolumn{2}{c}{$\mu_2$}\\
&& ($^{\circ} $) & $R_C$ & $B_C$&& $R_C$ & $B_C$& &\multicolumn{1}{c}{$R_C$} & \multicolumn{1}{c}{$B_C$}\\
\hline
\noalign{\smallskip}
CoRoT-1b & $4.72\pm0.17$ & $83.8\pm1.0$ & $0.1440\pm0.0028$ & $0.1471\pm0.0030$ && $0.58\pm0.31$ &  $0.34\pm0.27$ &&$-0.16\pm0.45$ & $0.38\pm0.38$ \\
CoRoT-2b & $6.76\pm0.09$ & $88.06\pm0.72$ & $0.1655\pm0.0015$ & $0.1665\pm0.0021$ && $0.38\pm0.10$ & $0.55\pm0.20$ & &$0.03\pm0.21$ & $0.06\pm0.33$ \\
CoRoT-4b & $16.60\pm1.05$ & $89.01\pm0.75$ & $0.1067\pm0.0025$& $0.1052\pm0.0041$ && $0.40\pm0.29$ &  $0.55\pm0.31$ &&$-0.01\pm0.41$ & $-0.03\pm0.44$ \\
CoRoT-5b & $9.44\pm0.77$ & $85.55\pm0.72$ & $0.1144\pm0.0030$ & $0.1104\pm0.0050$& & $0.45\pm0.30$  & $0.54\pm0.35$ & &$0.00\pm0.39$& $0.08\pm0.47$ \\
CoRoT-6b & $17.96\pm0.89$ & $89.04\pm0.61$ & $0.1152\pm0.0024$ & $0.1174\pm0.0031$ && $0.29\pm0.23$  & $0.51\pm0.29$ & &$0.20\pm0.41$& $0.00\pm0.45$ \\
CoRoT-7b & $5.40\pm1.10$ & $84.8\pm4.5$ & $0.0182\pm0.0015$& $0.0158\pm0.0088$& & $0.63\pm0.29$  & $0.45\pm0.35$ & &
$0.01\pm0.39$ &$0.00\pm0.38$ \\
CoRoT-8b & $18.3\pm2.7$ & $88.56\pm1.03$ & $0.0811\pm0.0034$& $0.0755\pm0.0072$ && $0.58\pm0.30$  & $0.50\pm0.32$ & &$-0.07\pm0.41$& $0.04\pm0.39$ \\
CoRoT-9b & $96.5\pm5.0$ & $89.84\pm0.12$ & $0.1144\pm0.0026$&  $0.1202\pm0.0042$ && $0.42\pm0.32$  & $0.42\pm0.34$ & &$-0.11\pm0.48$& $0.07\pm0.44$ \\
CoRoT-11b & $6.71\pm0.37$ & $82.86\pm0.62$ & $0.1009\pm0.0034$& $0.1031\pm0.0024$ && $0.44\pm0.34$  & $0.52\pm0.33$ & &$0.05\pm0.41$& $-0.04\pm0.38$ \\
\noalign{\smallskip}
CoRoT-3b & $8.3\pm1.3$ & $86.5\pm2.6$ & $0.0686\pm0.0030$& $0.0603\pm0.0080$ && $0.52\pm0.35$ & $0.52\pm0.34$ & &$-0.15\pm0.49$ & $0.00\pm0.39$ \\
\noalign{\smallskip}
\hline
\end{tabular}
\label{depths}
\end{table*}

Regarding the orbital parameters, all orbits were considered to be circular, except for CoRoT-5b ($e=0.07\pm0.06$) and 
CoRoT-9b ($e=0.07\pm0.05$), for which eccentricities
were left as free parameters but constrained around the values found in literature.
The resulting values of the inclination angle and of $a/R_s$ are all within the error bars of those given in the discovery papers.

We also investigated 
the possibility that the decision to fit simultaneously the limb-darkening parameters
could have affected the $R_p/R_s$ determinations.  We repeated the same fitting procedure keeping fixed the linear $\mu_1$ 
parameters to the median values (0.44 for $R_C$, 0.51 for $B_C$). We obtained the same values for the radius ratios, 
with negligible differences with respect to the values listed in Table \ref{depths}.

\begin{table} 
\centering
\caption{Comparison between fitted limb-darkening coefficients on the $W_C$ light curves 
and those estimated by Sing (2010) for the {\it CoRoT} range of sensitivity.}
\setlength{\tabcolsep}{5pt}
\begin{tabular}{ccccrc}
\hline
Planet & \multicolumn{2}{c}{$\mu_1$} && \multicolumn{2}{c}{$\mu_2$}\\
& $W_C$ & Sing& & \multicolumn{1}{c}{$W_C$} & Sing\\
\hline
\noalign{\smallskip}

CoRoT-1b & $0.51\pm0.27$ &  $0.37$ && $-0.03\pm0.40$ & $0.27$\\

CoRoT-2b & $0.43\pm0.10$ &  $0.46$ && $0.02\pm0.20$ & $0.22$\\

CoRoT-4b & $0.43\pm0.36$ &  $0.37$ && $0.04\pm0.52$ & $0.28$\\

CoRoT-5b & $0.37\pm0.31$ &  $0.37$ && $0.11\pm0.38$ & $0.27$\\

CoRoT-6b & $0.36\pm0.26$ &  $0.38$ && $0.09\pm0.43$ & $0.27$\\

CoRoT-7b & $0.48\pm0.35$ &  $0.53$ && $-0.02\pm0.49$ & $0.17$\\

CoRoT-8b & $0.52\pm0.31$ &  $0.59$ && $-0.03\pm0.40$ & $0.12$\\

CoRoT-9b & $0.54\pm0.25$ &  $0.46$ && $0.17\pm0.37$ & $0.22$\\

CoRoT-11b & $0.46\pm0.34$ &  $0.34$ && $0.02\pm0.40$ & $0.29$\\
\noalign{\smallskip}
CoRoT-3b & $0.44\pm0.34$ &  $0.32$ && $0.00\pm0.43$ & $0.31$\\
\noalign{\smallskip}
\hline
\end{tabular}
\label{ldwhite}
\end{table}

\subsection{The case of CoRoT-3b}

With its mass of 21.66 $M_J$ (Deleuil et al. 2008), CoRoT-3b was defined as an inhabitant of 
the `brown dwarf desert'. 
It could be a low-mass brown dwarf or the member of a new class of `superplanets', because it is on 
the edge between planets and stars. Therefore, we did not consider it as a bona-fide member
of our class of exoplanets, but the presence of two close stars 2.9 and 4.9 mag
fainter at distances of 5.6 and 5.3 arcsec, respectively (Deleuil et al. 2008),  
makes its analysis very challenging for the relevance of the contamination (Table~\ref{contaminants}).

\begin{figure} 
\begin{center}
\includegraphics[width=8.5cm]{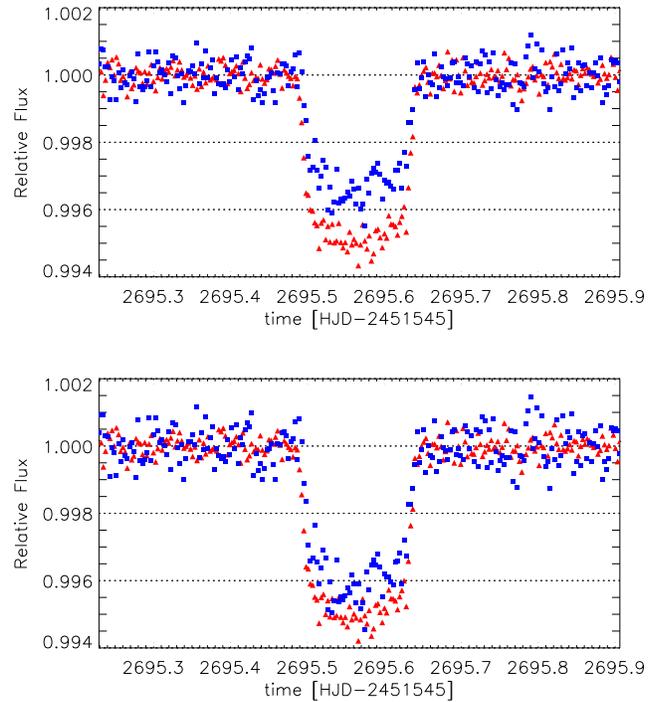}
\caption{Transits of CoRoT-3b in the $R_C$ (triangles) and
$B_C$ (squares) colours,
before {\it(top)} and after {\it(bottom)}  removing the
contamination of nearby stars.
Points refer to a binning of 0.003 d.
}
\label{corot3}
\end{center}
\end{figure}

The light curves of the original chromatic data, without applying any correction for the contaminant
stars, show a clear difference of 0.002 in relative flux in the depths of the transits (Fig.~\ref{corot3}, upper panel).  
After applying our method to remove
the contamination using the photometry listed in EXODAT, this  amount is halved (Fig.~\ref{corot3},
lower panel), confirming that  our procedure works in the right direction.
The resulting difference  between the radii in $R_C$ and $B_C$ is  still large ($\sim$12 per cent, 
see Table~\ref{depths}), but the scattered $B_C$ light curve and its quite large error bar makes the values of the radii 
compatible in this extremely contaminated case. 
In Fig.~\ref{corot3} a small `bump' is also notable at the bottom of the $B_C$ transit.
Due to the fact that the orbital period of CoRoT-3b is compatible with a synchronization with the 
stellar rotation (Mazeh \& Faigler 2010), we cannot exclude it to be an inhomogeneity on the surface of the F3V parent star. 
We also should not forget the specificity of the nature of
the CoRoT-3 system.   
For the sake of completeness, we calculated that  a missed bright contaminating star with $V$=16.0 and $B-V$=+0.3 
could equalize the depths of the transits when affecting the $B_C$ contamination up  to  35 per cent.
Undetected or undetectable contaminant stars, especially in the 
crowded field of the Galactic plane observed by {\it CoRoT}, constitute  a possible error source we cannot
quantify. However, at the moment, there is no evidence of a missed contaminating star around CoRoT-3
(Deleuil et al. 2008).

\section{Discussion}

\begin{figure}
\begin{center}
\includegraphics[width=8.5cm]{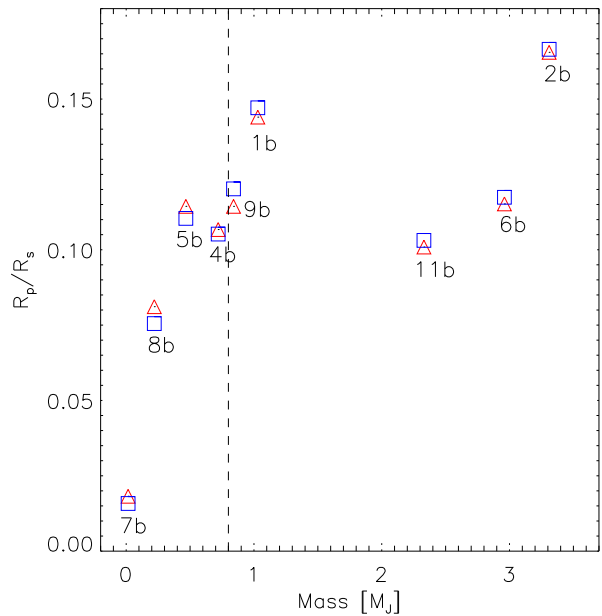}
\caption{Planet-to-star radius ratios of the {\it CoRoT} planets analysed in the $R_C$ (triangles) and $B_C$ (squares) colours. 
Error bars are of the order of the point dimension (except for CoRoT-7b which are about twice the point dimension).
}
\label{prova}
\end{center}
\end{figure}

\begin{figure} 
\begin{center}
\includegraphics[width=8.5cm]{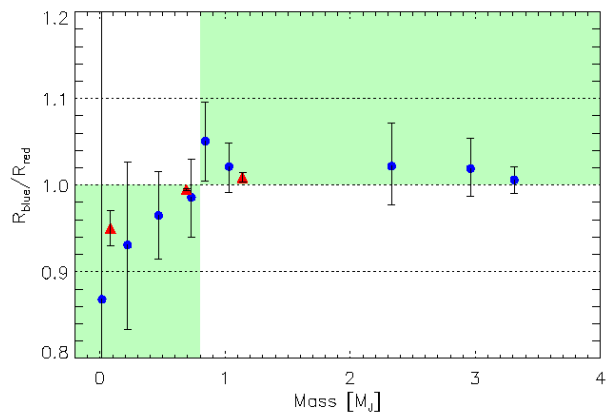}
\caption{Radius of the planet in the $B_C$ colour with respect to the radius in the $R_C$ colour. Blue circles refer
to our analysis of {\it CoRoT} planets. The huge error bars
of CoRoT-7b, not even completely inside the plot, are due to the very small depths of the transit with respect to the 
S/N, especially in the $B_C$ colour. 
Red triangles refer to literature data (cited in Section 3) taken in different optical bandpasses: $R_{blue}$ refers to the measurement with
the shorter wavelength, and $R_{red}$ to the measurement with the longer one.
}
\label{raggi}
\end{center}
\end{figure}

The planet-to-star radius ratios in the $R_C$ and $B_C$ colours 
(Table~\ref{depths})
provide the expected evidence that the transits are mostly achromatic.
In several cases, the ratio values
are far from equality, but the error bars recover it in all cases except   
for CoRoT-9b. Therefore, 
we looked at the whole set of planet-to-star radius 
ratios to verify if the ensemble analysis could provide some hints of 
departures from equality.
To do that we used 
the distance from the parent star,
the mass of the planet and the effective temperature of the planet.
We obtained the plot shown in Fig.~\ref{prova} when 
sorting the planet-to-star radius ratios in function of  the planet's mass. 
We can see that all the planets with a $B_C$ radius smaller than the $R_C$ radius
are confined in the $M<0.8~M_J$ region. We checked if such behaviour could depend from the contamination level,
but the plot did not change by varying the contamination factors up to 30 per cent of the value
calculated by using the procedure described in Section~2.1.
This could mean that the $R_C$ radius of the low-mass planets is larger than the $B_C$ radius, 
while for massive planets the radii are almost equal (Fig.~\ref{raggi}). The statistics
is very limited at the moment and the fact that all the planets with $M<0.8~M_J$ are
below the equality line could be fortuitous, also considering that they are those
with the bigger error bars.
For the sake of completeness, we verified if the behaviour sketched by Figs.~\ref{prova} and \ref{raggi} can reflect
some physical reason.
Tessenyi et al. (2012) stated that the height of a planet's atmosphere can be estimated to be $\Delta z=nH$, 
where typically $n\sim 5$ and $H$ is the atmospheric scaleheight.
Since $H = \frac{kT}{\mu g}$ \footnote{$k$ is the Boltzmann constant, 
$T$ the equilibrium temperature of the planet, $\mu$
the mean molecular mass of the atmosphere and $g$ the gravity acceleration.}, 
the smaller gravity of low-mass planets results in a larger atmospheric scaleheight, 
and this could enhance the atmospheric features in the photometric measurements
at different wavelengths. 
We also verified that the observed transit depth differences of some hundreds of ppm
are consistent with the effects of  planetary atmospheres.
In the specific case of the {\it CoRoT} planets, we obtained values of
$\Delta z$ in the range 0.5-5 per cent of the radius of the planet (even more for CoRoT-7b).
These values of the atmospheric height correspond to upper limits on transit depth differences 
due to an atmospheric absorption of 100-1300 ppm.

We searched for other transiting planets with transit depth measurements in multiple optical bandpasses.
The characterization of the planetary atmosphere of HD 189733b ($M=1.138~M_J$) supplied evidences  
of the presence of high-altitude atmospheric haze detected in the visible 
(Pont et at. 2008; Lecavelier Des Etangs et al. 2008;
Sing et al. 2011) and probably extended to the infrared (Gibson et al. 2012).
The transmission spectrum is in good agreement with that of Rayleigh scattering 
and the planetary radius results to be slightly  smaller at long wavelengths, in agreement with
the results obtained from {\it CoRoT} photometry (Fig.~\ref{raggi}). 
The  mass of HD~209458b ($M=0.69~M_J$) is too close to the value $0.8~M_J$ to provide a useful 
check on 
the dependence of the planet-to-star radius ratios on the mass, but we note that Knutson et al. (2007)
obtained  different radius values in function of wavelength in the full range from 293 to 1019~nm.
Theoretical transmission spectra support such transit depth variations 
(Gillon et al. 2012, see their fig.~7).
With a mass of only $0.081~M_J$, HAT-P-11b could provide a significant test. 
Deming et al. (2011) pointed out that the radius ratio is 6 per cent larger 
in the $J$ band than in the {\it Kepler} photometry. They are not able to provide an
explanation for such a large discrepancy and suggest a variable contribution of the water absorption in
the Earth's atmosphere during the observations.

\section{Conclusions}
We provided a methodological approach to extract the chromatic information contained
in the light curves of 10 transiting planets discovered by {\it CoRoT}. We satisfactorily
removed the contaminating light coming from 
poorly known stars inside the pre-defined aperture masks.
This step is of extreme importance for a coherent analysis of transits, as third light 
can affect in different percentages the chromatic light curves and thus bring to incongruous comparisons
between the different colours.
When comparing transits of the same planet observed with different instruments (or at different wavelengths), 
an analysis of the contamination effect is fundamental,
since third light could lead to different and erroneous transit depth (and planetary radius) estimations.

Then we detrended and cleaned
the data producing a homogenous set of light curves in each of the {\it CoRoT} colours.
The analysis of the {\it CoRoT} chromatic light curves
confirms that all the transit depths for the planets studied in two
different colours are nearly equal.
However, we could put in evidence an interesting behaviour of the
planet-to-star radius ratios of the transits in $B_C$ and $R_C$ colours when they are plotted versus the
mass of the planets. We suggest the possibility of observing departures from transit depth equality for planets 
having $M<0.8~M_J$. 
This hypothesis is still  rather speculative at the moment, but is not groundless, roughly reflecting
the relevance of the atmospheric scale height in low-mass planets. 
The inclusion of the
Na~D lines in the range of the $R_C$ colour (L{\'e}ger et al. 2009) could also play a role
when considering that Sing et al. (2008a, 2008b) report on the very strong 
absorption of the Na~D lines in the high atmosphere
of HD~209458b, around 589~nm. Of course,
the results would have been more compelling if the {\it CoRoT} photometric
system was a standard or a well-calibrated one. 

The careful evaluation of the chromatic effect could
result in a fertile exercise in the case of low-mass planets. 
High-precision multicolour photometry of transits could provide an efficient tool
to investigate the upper atmosphere of exoplanets,  taking into account the intrinsic difficulties
inherent in accessing these thin layers in other worlds' atmospheres.

\section*{Acknowledgments}
The authors wish to thank D.~Pollacco for useful comments on a first
draft of the manuscript, J.~Vialle for checking the English form, and
the anonymous referee  for the suggestions aimed at improving the
presentation of the results.
FB acknowledges financial support from the ASI-Universit\`a di Padova Contract I/044/10/0. 
EP acknowledges financial support from the PRIN-INAF 2010 {\it 
Asteroseismology: looking inside the stars}.

\label{lastpage}

\end{document}